\documentclass[twocolumn,showpacs,preprintnumbers]{revtex4}
\usepackage{amsmath,amsfonts,amssymb,graphics,graphicx,epsfig,color,times,bbm}

\begin{document}

\preprint{}

\title{Irreversible photon transfer in an ensemble of $\Lambda$-type atoms and photon diode}
\author{Gor Nikoghosyan$^{1,2}$ and Michael Fleischhauer$^{1}$}
\affiliation{Dept. of Physics and research center OPTIMAS,
University of Kaiserslautern, D-67663 Kaiserslautern, Germany$^{1}$\\
Institute of Physical Research, 378410, Ashtarak-2, Armenia$^{2}$}
\pacs{42.50.Dv, 42.25.Kb, 03.65.Yz, 03.65.Xp}
\begin{abstract}
We show that a pair of quantized cavity modes interacting with a spectrally
broadened ensemble of $\Lambda $-type atoms is analogous to an ensemble
 of two level systems
coupled to a bosonic reservoir. This provides the possibility for an irreversible photon transfer
between photon modes. The density of states as well as the quantum state of the reservoir can  be engineered 
allowing the observation of effects such as the quantum Zeno- and anti-Zeno effect, the destructive interference of
decay channels and the decay in a squeezed vacuum. As a particular application we discuss 
a photon diode, i.e. a device which directs a single photon from anyone of two input ports to a common output port.
\end{abstract}
\maketitle

In the present paper we propose a photonic analogue of the irreversible decay of 
an ensemble of two-level systems coupled to a bosonic reservoir \cite{Zoller}.
In particular we consider a pair of quantized cavity modes
interacting with a spectrally
broadened ensemble of $\Lambda $-type atoms. 
The two cavity modes replace the collective states of the ensemble of two-level systems and the $\Lambda$-type atoms 
form the modes of the reservoir. 
In contrast e.g. to the quantized radiation field as a reservoir, the atomic ensemble can 
be easily modified and controlled dynamically, which can be used for reservoir engineering \cite{Wineland}. E.g. the density of states of 
the reservoir can be tailored by application of inhomogeneous magnetic or electric fields
and thus it should be possible to implement e.g. the quantum Zeno \cite{Sudarshan} and anti-Zeno \cite{Kurizki1, Kurizki2, Facchi} effects, which
are otherwize difficult to realize.
Moreover the reservoir of $\Lambda$ atoms can be prepared in different initial quantum states.
E.g. coherent ensemble states can be created by
using electromagnetically induced transparency \cite{Fleischhauer}
or methods of adiabatic population transfer \cite{Bergmann}. Also non-classical states 
can be prepared \cite{Polzik1, Polzik2, Polzik3, Oberthaler} which can be used to simulate a squeezed vacuum reservoir \cite{Gardiner1, Gardiner2, Zubairy}.
If the atomic ensemble is prepared in only one internal state, serving as the
vacuum of reservoir excitations, the analogue
of spontaneous decay can be observed, where the photons of one cavity mode
are transferred irreversibly, i.e. non-unitarily, to the second mode. 
This effect can have a variety of applications; e.g.  creation of new
quantum states,  transfer of photons of optical frequency to the microwave domain
and vice versa, or a photon diode, i.e. a device 
where a single photon injected into anyone of two inputs ports leads to a single photon
emission from the same output port.
The system considered here can easily be constructed with current technology and 
is available in several labs. In a number of labs strong coupling of a cavity mode with 
a Bose-Einstein condensate of atoms \cite{Esslinger1, Reichel-Nature-2007, Stamper-Kurn-2007} or a cold atomic cloud \cite{Aspelmeyer, Jaehne} 
was achieved. As the two modes one can use orthogonal polarizations of the same 
frequency and the required spectral broadening of the atomic ensemble can be achieved e.g. by application of an inhomogeneous magnetic field.
\begin{figure}
  \epsfig{file=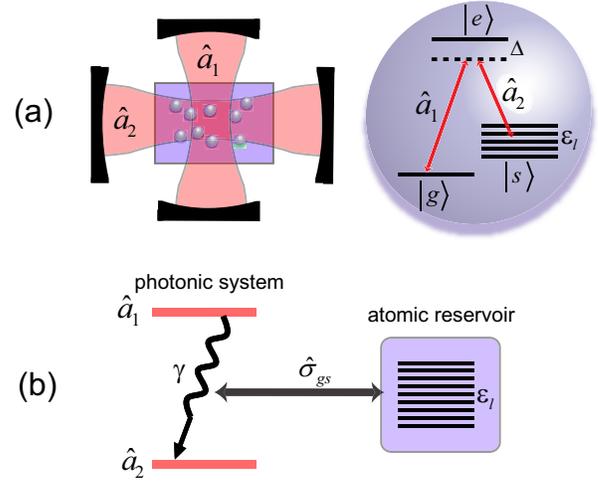,width=0.9\columnwidth}
    \caption{ (Color online) (a)  Schematic setup: Two quantized cavity modes $\hat a_1$ and $\hat a_2$
interact with an ensemble of three-level $\Lambda$ type atoms with inhomogeneously broadened
two-photon transition $|g\rangle -|s\rangle$.
(b) For a large number of three-level atoms and a sufficiently large spectral width of the
Raman transition the system resembles a collection of two-level
systems coupled to a bosonic reservoir
}
  \label{fig:Couplings}
 \end{figure}

Let us 
consider the interaction of two light modes
described by the annihilation operators $\hat a_1$ and $\hat a_2$
with an ensemble of three level $\Lambda $-atoms (see fig.\ref{fig:Couplings}).
$\hat a_1$ and $\hat a_2$ couple the ground-state $|g\rangle$, respectively a
meta-stable lower state $|s\rangle$ to a common excited state $|e\rangle$ in a
Raman transition. The two-photon transition between states $|g\rangle$ 
and $|s\rangle$ is assumed to be inhomogeneoulsy broadened, as indicated in
fig.\ref{fig:Couplings}. For simplicity we assume a discrete spectrum consisting of $f$
energy levels. In each spectral class there are $N$ atoms, so that the total number of atoms
is $f N$.

To describe the quantum
properties of the medium, we use collective atomic operators for each spectral component
$\hat\sigma _{ij}^{l}=\frac{1}{\sqrt{N}}
\sum\limits_{k=1}^{N}\left\vert i\right\rangle _{kk}^{ll}\left\langle
j\right\vert $, where $i,j\in {e,g,s}$, and $k$ labels the atom.
The dynamics of the system is then governed by the Hamiltonian
\begin{eqnarray}
\hat{H} &=&\hbar \sqrt{N}\sum\limits_{l=1}^{f}\Bigl[\Delta \hat{\sigma}%
_{ee}^{l}+\varepsilon_{l}\hat{\sigma}_{ss}^{l}+  \label{hamiltonian} \\
&&+\left( g_{1}\hat{\sigma}_{eg}^{l}\hat{a}_{1}+g_{2}\hat{\sigma}%
_{es}^{l}\hat{a}_{2}+H.c.\right) \Bigr]  \notag,
\end{eqnarray}
where  $\Delta $ is the one-photon detuning of the cavity field
1, and $\varepsilon_{l}$ is the two-photon
detuning corresponding to the $l$th spectral class. $g_{1,2}$ are the coupling
strength of both modes, which are assumed to be the same for all spectral components 
for simplicity. The factor $\sqrt{N}$ is due to the collective coupling of 
the atoms of each spectral class to the cavity modes \cite{Tavis, Dicke}.
If the vacuum Rabi frequencies $g_{1,2}$ and the two-photon
detunings $\varepsilon_{l}$ are significantly smaller than $\Delta $,
we can adiabatically eliminate the upper level $\left\vert e\right\rangle $
by standard techniques. 
Let us further assume that the atomic ensemble is prepared in a state which
is close to the collective ground state and that the total number of photons
is much less than the total number of atoms. Then the population of  state $|s\rangle$ remains small 
and the atomic operators 
$\hat{\sigma}_{gs}^{l}$ and $\hat{\sigma}_{sg}^{l}$ obey aproximately bosonic
commutation relations $\left[ \hat{\sigma}_{sg}^{l},\hat{\sigma}_{sg}^{p}\right] \approx 0$,
$\left[ \hat{\sigma}_{gs}^{l},\hat{\sigma}_{sg}^{p}\right]\approx \delta _{l,p}$, i.e. we can set
$\hat{\sigma}_{gs}^{l}=\hat \beta_l$, $\hat{\sigma}_{sg}^{l}=\hat \beta_l^\dagger$. Thus passing to
an interaction picture we arrive at an effective Hamiltonian
\begin{equation}
\hat H_{\rm eff} = -\hbar \sqrt{N} \sum_{l=1}^f \Bigl\{\eta\hat\beta_l \hat a_1^\dagger \hat a_2 \, {\rm e}^{i\omega_l t}
+h.c.\Bigr\},
\end{equation}
where $\eta= (g_1)^* g_2/ \Delta$ and $\omega_l=\varepsilon_l+|g_1|^2/\Delta$ is an effective detuning containing ac-Stark shift contributions.

$\hat H_{\rm eff}$ is similar to the Hamiltonian that
describes the interaction of an ensemble of two level systems consisting of states
$|1\rangle$ and $|2\rangle$ with a reservoir
of bosonic modes $\hat \beta_l$. $\hat a_1^\dagger \hat a_2$ destroys an atom
in state $|2\rangle$ in the ensemble and creates an atom in state $|1\rangle$.

The dynamics of a two-level system that interacts with a
reservoir becomes irreversible when the number of modes of the bath 
tends to infinity. A well-known consequence of this sort of dynamics is spontaneous
decay and a similar dynamics can be obtained here for photons.
To see this we derive an effective equation of motion for the photon modes by
tracing out the atomic degrees of freedom using standard second order perturbation 
theory. This yields for the density operator $\rho$ of the photon modes
${\dot{\hat{\rho}}}(t) = -\int\limits_{0}^{t}{\rm d}\tau
g( t-\tau)\Bigl[ \hat{a}_{1}^\dagger(t)
\hat{a}_{2}(t) \hat{a}_{1}(\tau) \hat{a}_{2}^\dagger(\tau) \hat{\rho}(\tau)
 - \hat{a}_{1}(\tau) \hat{a}_{2}^\dagger(\tau)\hat{\rho}(\tau)
\hat{a}_{1}^\dagger(t)\hat{a}_{2}(t)\Bigr] +h.c. $
where $g\left( t\right) =N\sum_{l=1}^{f}\left\vert
\eta\right\vert ^{2}e^{i\omega _{l}t}$.
The behavior of the system depends on the reservoir response
function $g\left( t\right) $ which is determined by the effective light--field
coupling constants $\eta$ and the resonance frequencies
$\omega _{l}$ of the atomic ensemble \cite{Breuer}. We assume $f$ to be large enough so that the recurrence time is large and can be disregarded. For simlicity we also assume that the ensemble has equidistant spectral
lines with $\omega_{l}=\varepsilon
_{\max }( 2l-1-f) /{f}$, where $2\varepsilon _{\max }$ is the
spectral width of the atomic ensemble. 
Under the mentioned conditions, $g(t)$ can be approximated as a delta function $g\left(
t\right) \approx \pi ({Nf}/{\varepsilon _{\max }})\delta \left( t\right)
\left\vert\eta\right\vert ^{2}$, which constitutes the Born-Markov approximation. 
In this limit we obtain
\begin{equation}
\frac{\partial {\hat{\rho}}}{\partial t}=-\frac{\gamma}{2}\left( \hat{a}_{1}^\dagger\hat{a}%
_{1}\hat{a}_{2}\hat{a}_{2}^\dagger{\hat{\rho}}+{\hat{\rho}}\hat{a}_{1}^\dagger\hat{a}_{1}%
\hat{a}_{2}\hat{a}_{2}^\dagger-2\hat{a}_{1}\hat{a}_{2}^\dagger{\hat{\rho}}\hat{a}%
_{1}^\dagger\hat{a}_{2}\right)   \label{bloch},
\end{equation}
with $\gamma ={\pi N f \left\vert\eta\right\vert ^{2}}/\left({\varepsilon _{\max }}\right)$.

The Lindblad-equation (\ref{bloch}) describes the irreversible transfer of
excitations from mode 1 to 2 with rate $\gamma $. If there are many photons
in the system the decay will be enhanced due to stimulated Raman emission
into mode 2. The atomic analogue of this process is the collective decay of
atoms (superfluorescence) \cite{Dicke}.

\paragraph*{(i) irreversible photon transfer}
Let us first consider the irreversible transfer of photons from mode 1 to
mode 2 for the case when mode 2 is initially not populated and the initial
state of the system is pure, i.e. $\hat\rho_0 =\vert \phi_0\rangle\langle\phi_0\vert$
with $\left\vert \phi_0 \right\rangle
=\sum\limits_{n}\alpha _{n}\left\vert n\right\rangle _{1}\left\vert
0\right\rangle _{2}$. where $\left\vert
n\right\rangle _{1}\left\vert m\right\rangle _{2}$ are Fock states of 
modes 1 and 2 respectively. This state will evolve into the mixed state
$\hat{\rho}_{\rm fin}=\sum\limits_{n}\left\vert \alpha _{n}\right\vert
^{2}|0\rangle_{11}\langle 0\vert\otimes\vert n\rangle_{22}\langle n\vert$.
It is pure only if mode 1 is initially in a single Fock
state.

More interesting is the general case of an initially mixed state of both modes
$
\hat{\rho}_{0}=\sum\limits_{{m,m^\prime },{n,n^\prime}}
A_{m,m^\prime}^{n,n^\prime } \left\vert n\right\rangle _{11}\left\langle n^\prime\right\vert
\otimes \left\vert m\right\rangle _{22}\left\langle m^\prime\right\vert
\label{purification1}
$
In this case the final state reads
$
\hat{\rho}_{\rm fin} =\sum\limits_{p,p^{\prime }}B_{p,p^{\prime }}|0\rangle_{11}\langle 0\vert
\otimes \left\vert p\right\rangle _{22}\left\langle p^{\prime }\right\vert
 \label{purification2}
$
where $B_{p,p^{\prime }} =\sum\limits_{q=0}
A_{p-q,p^{\prime }-q}^{q,q}$. This state is a pure state $\left\vert \phi
\right\rangle _{2}=\sum_m\beta _{m}\left\vert m\right\rangle
_{2}$ if $B_{p,p^{\prime }} =\beta_p^* \beta_{p^\prime}$.
This implies that all initial states with
$A_{n,n^{\prime }}^{m,m^{\prime }}=\beta _{m+n}\beta _{m^{\prime
}+n}^{\ast }\alpha _{n}\delta _{n,n^{\prime }}$, where $\alpha _{n}$ is an
arbitrary function of $n$, evolve into a pure state.
E.g., assume that is completely mixed and contains
only one photon $\hat{\rho}_{0}=\frac{1}{2}\left\vert 1\right\rangle _{11}
\left\langle 1\right\vert\otimes
\left\vert
0\right\rangle _{22}\left\langle 0\right\vert+\frac{1}{2}\left\vert 0\right\rangle _{11}\left\langle 0\right\vert 
\otimes\left\vert 1\right\rangle
_{22}\left\langle 1\right\vert $. Then the
final state will be a pure single photon Fock state $\hat{\rho}%
_{\rm fin}=\vert 0\rangle_{11}\langle 0 \vert\otimes\left\vert 1\right\rangle _{22}\left\langle 1\right\vert $.
I.e. an initially mixed states of light can be purified with 
\textit{conservation } of photon number. Another related example is the case when 
initially we have Fock states $\hat{%
\rho}_{0}=\left\vert n\right\rangle _{11}\left\langle n\right\vert \otimes
\left\vert m\right\rangle
_{22}\left\langle m\right\vert $ in both
modes.  The final state is again a pure Fock state $\hat{\rho}%
_{\rm fin}=\vert 0\rangle_{11}\langle 0 \vert\otimes
\left\vert n+m\right\rangle _{22}\left\langle n+m\right\vert \,$, that
contains the sum of initial photon numbers. The latter process can be realized
also by unitary operations, but for that it is necessary to know the exact number of
photons in each mode \cite{Haroche, Walther1}, while the process which we
describe is irreversible and can be realized without any information
about the initial photon numbers in each mode. It is enough to know, that
the initial states are Fock states.

\paragraph*{(ii) reservoir engineering}

As originally formulated by Mishra and Sudarshan a decaying quantum
system that is continously observed in a specific state does not
decay, which they called the quantum Zeno effect
\cite{Sudarshan}.
 In practice a continuous observation is approximated by a periodic sequence of
measurements. In order to observe the decay suppression the period of
the repeated projections has to be shorter than a characteristic time
determined by the spectral structure of the reservoir coupling. 
The latter makes the observation of the effect in the decay of a two-level atom
to the free-space electromagnetic vacuum rather difficult. In this case
the spectrum of the reservoir coupling is flat and 
the frequency of measurements has to be comparable to the transition frequency
in order to see the quantum Zeno suppression of decay.
If the reservoir spectrum is structured also the opposite
effect, called anti-Zeno effect is possible \cite{Kurizki1, Facchi}. The periodic interaction 
between system and measurement device shifts the effective resonance 
frequency of the system and moves it to a different part of the
reservoir spectrum. This can 
increase (decrease) the system--reservoir coupling, and thus lead to an  
increase (decrease) of the decay rate.
Similar effects can be observed by keeping the resonant frequency of 
the system unperturbed but shifting the spectrum of the reservoir. The 
latter can be realized by the measurement of the reservoir \cite{Kurizki2, Facchi}.
In our system the reservoir is atomic thus one can easily tailor the 
reservoir spectrum and measure it by application of an 
electromagnetic field. Depending on the reservoir response function $f(t)$ 
the measurement will either accelerate (anti-Zeno effect) or slow 
down (Zeno effect) the spontaneous photon transfer from one mode into 
another mode.

Another potential application of reservoir engineering is the possibility to observe 
destructive interference of decay channels. The radiative decay
in three-level $V$-type systems with nearly degenerate dipole transitions
can exhibit destructive interference and thus suppression of decay if the
dipole moments of the transitions from the two excited states to the common
ground state are parallel \cite{Scully}. Unfortunately in atomic systems
degenerate transitions with parallel dipole moments do not exist. 
Generalizing the present discussion to multi-mode photonic systems provides however
a means to observe such a destructive interference of decay channels. To this end consider 
a system where in addition to mode $\hat a_1$ a second cavity mode $\hat a_1^\prime$, e.g.
of orthogonal polarization, couples the same atomic transition $|g\rangle - |e\rangle$.
In this case the ``excited state '' $\hat a_1$ in Fig.\ref{fig:Couplings} b is replaced by two
``excited states'' $\hat a_1$ and $\hat a_1^\prime $ coupling to the common atomic reservoir.
Then in the Lindblad eq.(\ref{bloch}) $\hat a_1$ is replaced by $(\hat a_1+\hat a_1^\prime)$.
As a consequence there are non-decaying states, as e.g. the anti-symmetric single-photon excitation
$(|0\rangle|1\rangle -|1\rangle|0\rangle)/\sqrt{2}$.

Furthermore it is also possible to prepare the reservoir in a certain quantum state,
e.g. in multi-mode squeezed states \cite{Polzik1, Polzik2, Polzik3, Oberthaler}. With this it should be possible to study the decay
in a squeezed vacuum \cite{Gardiner1, Gardiner2, Zubairy}. The latter has been proposed
and analyzed theoretically for atoms coupled to a squeezed reservoir of radiation modes. 
An experimental verification is however extremely difficult due to the requirement of a 
broad-band and isotropic squeezed-vacuum radiation field.

\paragraph*{(iii) photon diode}
In the last part of this paper we discuss an interesting application
of the irreversible photon transfer, the realization of a diode for photons, i.e. a four-port device
where single photon pulses injected into any of the two input ports will be
directed to the same output port.  To model the input--output processes we
introduce a continuum of free-space modes with field operators $\hat{b}_{1q}$
and $\hat{b}_{2q}$ which are coupled to the cavity modes $\hat{a}_{1}$ and $\hat{a}%
_{2}$ respectively. For simplicity we assume that the coupling constants $%
\kappa _{1}$ and $\kappa _{2}$ are the same for all relevant modes. The corresponding
interaction is described by the following Hamiltonians $(m=1,2)$
\begin{equation*}
\hat{V}_{m}=\hbar\kappa _{m}\sum\limits_{q}\left( \hat{a}_{m}^\dagger\hat{b}%
_{mq}+h.c.\right) +\sum\limits_{q}\hbar\Delta _{m}^{q}\hat{b}_{mq}^\dagger\hat{b}%
_{mq}.
\end{equation*}
Here $\Delta _{m}^{q}$ are the detunings of free field modes from the cavity
resonance. Let us consider input fields in a single-photon state $\left\vert
\psi _{in}\right\rangle _{m}=\sum\limits_{q}P_{q}^{in}\left( t\right) \hat{b%
}_{mq}^\dagger\left\vert 0\right\rangle $.
All properties of the fields
are then described by the single-photon wavefunction 
 $\Phi _{in}^{m}\left( z,t\right)
=\sum\limits_{q}\left\langle 0\right\vert \hat{b}_{mq}e^{iqz}\left\vert
\psi _{in}\right\rangle _{m}$.
Since all atoms of the atomic ensemble are initially in the ground
state, the field in state $\left\vert \psi _{in}\right\rangle _{2}$ sees an empty cavity 
and will be reflected from it with some time delay corresponding to the bare cavity
ring down time. In the following we want to prove that $\left\vert \psi
_{in}\right\rangle _{1}$ is transferred to $\left\vert \psi
_{out}\right\rangle _{2}$, i.e. to a state where the excitation is in
the orthogonal output channel.
In  general the state of the system can be written in the following form%
\begin{eqnarray*}
\left\vert \psi \left( t\right) \right\rangle &=&\Bigl(
\sum\limits_{q}P_{q}\left( t\right) \hat{b}_{1q}^\dagger+Q\left( t\right)
\hat a_{1}^\dagger +\sum\limits_{l=1}^{f}R_{l}\left(
t\right) \hat a_{2}^\dagger\hat \beta
_{l}^{\dagger}+\\
&&\quad +\sum\limits_{l=1}^{f}\sum\limits_{q}S_{ql}\left( t\right) \hat{b}
_{2q}^\dagger\hat\beta
_{l}^{\dagger}\Bigr) \left\vert {\bf gs}\right\rangle,
\end{eqnarray*}
with $|{\bf gs}\rangle$ denoting the ground state of the atomic ensemble, where
all atoms are in state $|g\rangle$.
Since initially only mode 1 is excited and all atoms are in the ground state
$Q\left(t\right) =R_{l}\left( t\right) =S_{ql}\left( t\right) =0$ if $t<0$,
where $t=0$ is the beginning of interaction. The evolution of the system
is described by the Schr\"{o}dinger equation
\begin{eqnarray}
\overset{.}{P}_{q}\left( t\right)  &=&-i\Delta _{1}^{q}P_{q}\left( t\right)
-i\kappa _{1}Q\left( t\right),   \label{sch1} \\
\overset{.}{Q}\left( t\right)  &=&-i\kappa _{1}\sum\limits_{q}P_{q}\left(
t\right) + i\eta\sqrt{N}\sum\limits_{l=1}^{f} e^{i\omega _{l}t}R_{l}\left(
t\right),   \label{sch2} \\
\overset{.}{R}_{l}\left( t\right)  &=&i\eta^*\sqrt{N} e^{-i\omega
_{l}t}Q\left( t\right) -i\kappa _{2}\sum\limits_{q}S_{ql}\left( t\right),
\label{sch3} \\
\overset{.}{S}_{ql}\left( t\right)  &=&-i\Delta _{2}^{q}S_{ql}\left(
t\right) -i\kappa _{2}R_{l}\left( t\right).   \label{sch4}
\end{eqnarray}
Substituting the formal solution of eq.(\ref{sch4}) into (\ref{sch3}) and assuming the Markov limit yields
\begin{equation}
\overset{.}{R}_{l}\left( t\right) =i\eta^*\sqrt{N} e^{-i\omega
_{l}t}Q\left( t\right) -\frac{\gamma _{2}}{2}R_{l}\left( t\right),
\label{sch51}
\end{equation}
where $\gamma _{2}={\kappa _{2}^{2}L}/{c}$ is the cavity loss rate of
mode 2 and $L$ is the quantization length of $\hat{b}$ modes.
(Note that $\kappa_m\sim1/\sqrt{L}$, so that the dependence on $L$ drops.)
Furthermore  substituting the formal solutions of eqs. (\ref{sch51}) and (\ref{sch1}) 
into (\ref{sch2}) we find again using the Markov approximation
\begin{equation}
\overset{.}{Q}\left( t\right) =-\frac{\left(\gamma +\gamma _{1}\right) }{2}%
Q\left( t\right) -i\kappa _{1}\sum\limits_{q}P_{q}\left( 0\right)
e^{-i\Delta _{1}^{q}t },  \label{sch6}
\end{equation}
where we have used the photon decay rate $\gamma $ and introduced the cavity loss
rate of mode 1, $\gamma _{1}={\kappa _{1}^{2}L}/{c}$. 
Upon integrating eq. (\ref{sch6}) we finally find the input-output relation for port 1,
i.e. for the modes $\hat{b}_{1}^{q}$.
\begin{equation}
\Phi _{out}^{1}\left( t\right) =\Phi _{in}^{1}\left( t\right) - \gamma
_{1} F(t),
\label{sch8}
\end{equation}
where $ F(t) = \int\limits_{{0}}^{t}\mathrm{d}\tau \, \Phi _{in}^{1}\left( \tau \right)
e^{-\frac{\left(\gamma +\gamma _{1}\right) }{2}\left( t-\tau \right) }$. In order to achieve a maximum transfer of free-field photons into the cavity,
the outgoing component should be minimized. According to eq.(\ref{sch8}) this
can be realized by requiring impedance matching. If $\gamma
_{1}=\gamma $ and the pulse is much longer than the relaxation rates, 
the two terms on the right hand side of eq.(\ref{sch8}) cancel each other and
there is no output  into the modes $\hat{b}_{1q}$.

Due to the dissipative nature of the coupling between the two cavity modes
the output field will be in general in a mixed state when tracing out the
degrees of freedom of the atomic ensemble. Only if the input is an eigenstate of the total
excitation number, the cavity output can be in a pure state.
We will show now that
in this case the final state indeed factorizes into a single photon distributed
over many modes $\hat b_{2q}$ and a single collective excitations in the 
atomic ensemble.
Following similar steps than above we find  for the output wave function for the modes $\hat{b}_{2q}$ when the
atomic excitation is in mode $n$, $\Phi_{\rm out}^{2,l}
\equiv \sum_q\langle 0|\langle 1_l|\hat b_{2 q} |\psi\rangle_{\rm out}$
where $|1_l\rangle$ denotes a single excitation in the $l$th spectral
class of the atomic ensemble, reads
\begin{eqnarray*}
&&\Phi _{out}^{2,l}\left( t\right) 
= -i\frac{\kappa _{1}}{\kappa _{2}}\gamma_{2}\eta^*\sqrt{N}
\int\limits_{{0}}^{t}\mathrm{d}\tau _{1}\, e^{-i\omega _{l}\tau_{1}}
e^{-\frac{\gamma _{2}}{2}\left( t-\tau _{1}\right)} F(\tau_{1}).
\end{eqnarray*}
Thus the probability of having a photon in output port 2, obtained
by tracing out the atomic part, reads
\begin{eqnarray*}
\rho _{out}\left( t\right)  &=&\sum\limits_{l}\left\vert\Phi _{out}^{2,l}\right\vert^2
 =\gamma _{1}\gamma _{2}\gamma \int\limits_{{0}}^{t}\mathrm{d}\tau
_{1}e^{-\gamma _{2}\left( t-\tau _{1}\right) }\left\vert F(\tau_{1})\right\vert^2.
\end{eqnarray*}
If the cavity decay rate of mode 2, $\gamma _{2}$,  is much larger than the relaxation rates 
$\gamma $ and $\gamma _{1}$, the integral over $\tau _{1}$ can easily be evaluated. 
In this case $\rho _{out}$ can be expressed as a product of two single photon
wavepackets $\rho _{out}\left( t\right) =\left\vert \Phi _{out}^{2}\left(
t\right) \right\vert ^{2}$, where
\begin{equation*}
\Phi _{out}^{2}\left( t\right) =\sqrt{\gamma _{1}\gamma }F(t).
\end{equation*}
Thus indeed a single-photon input wavepacket 
results into an asymptotic final state which is a product of
a collective atomic excitation and a single-photon wavepacket in
the output mode 2. It should be noted that since the physical mechanism
that leads to the diode function is irreversible, input superposition states
will in general not be mapped to pure output states. 
 \begin{figure}
  \epsfig{file=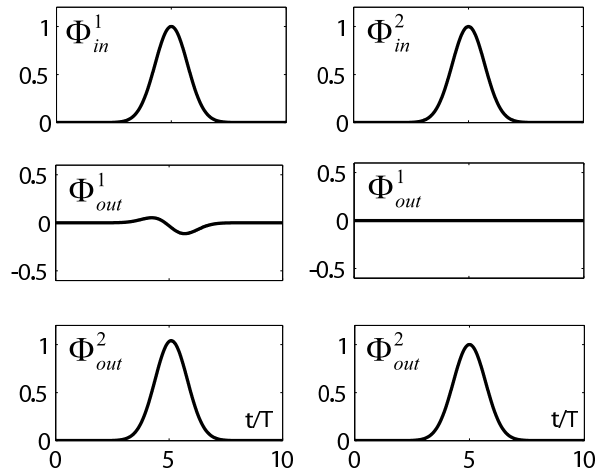,width=0.9\columnwidth}
    \caption{ {Left (right) column corresponds to a single photon input in chanel 1 (2). The photon 
injected into either of the two input modes is transfered to the same output port 2. The amplitude of the field in output channel 1 is negligible.}
}
  \label{fig:diode-example}
 \end{figure}
In fig.\ref{fig:diode-example} we have illustrated the performance of the
photon diode. Shown are two cases with a single-photon wavepacket
injected into input port 1 (modes $\hat b_{1q}$)  and input port 2 (modes
$\hat b_{2q}$). In both cases the output in modes 1 is negligible, while
the output in modes 2 is a nearly perfect single-photon wavepacket.

In conclusion we have shown that a two-mode system interacting with a
spectrally broadened ensemble of $\Lambda$-type atoms behaves as 
a collection of two-level systems interacting with a bosonic reservoir. 
The analogy between these two systems
allows the observation and simulations of several interesting phenomena of 
dissipative processes in engineered reservoirs.
In particular we have shown that similarly to spontaneous decay of
atoms one can irreversibly transfer photons from one mode to another.
The possibility to tailor the reservoir spectrum and to prepare 
collective quantum states of the ensemble of $\Lambda$ atoms 
can be used to observe the quantum Zeno- and anti-Zeno effects, to
study the decay in a squeezed reservoir and to observe destructive interference
of spontaneous decay channels. As a particular application we have discussed
in more detail a photon diode, i.e. a device where a single-photon input in any
of the two input ports is always emitted in only one of the two outputs. Besides being
intersting in its own right the diode may be used for the implemention of a classical
logical OR. 

G.N. thanks Jyrki Piilo for helpful discussion. The work of G.N. was supported by Alexander von Humboldt Foundation.

\end{document}